\documentclass[
aps,
prl,
10pt,
amssymb,
amsmath,
superscriptaddress,
tightenlines,
twocolumn,
notitlepage,
] {revtex4}
\usepackage{graphicx}
\usepackage[colorlinks, linkcolor = blue, citecolor = blue, urlcolor=blue, pdfborder={0 0 0 [0 0]},bookmarks=false]{hyperref}
\usepackage{bm}
\usepackage{physics}
\usepackage{amssymb}
\usepackage{amsmath}
\usepackage[dvipsnames]{xcolor}
\usepackage{lipsum}

\newcommand{\be}{\begin{equation}}
\newcommand{\ee}{\end{equation}}

\newcommand{\la}{\left \langle}
\newcommand{\ra}{\right \rangle}

\begin{document}

\title{Pseudogap metal and magnetization plateau from doping moir\'e Mott insulator}
\author{Yang Zhang}
\affiliation{Department of Physics, Massachusetts Institute of Technology, Cambridge, MA 02139, USA }
\author{Liang Fu}
\affiliation{Department of Physics, Massachusetts Institute of Technology, Cambridge, MA 02139, USA }

\begin{abstract}
The problem of doping Mott insulators is of fundamental importance and long-standing interest in the study  of strongly correlated electron systems. The advent of semiconductor based moir\'e materials opens a new ground for simulating the Hubbard model on the triangular lattice and exploring the rich phase diagram of doped Mott insulators  as a function of doping and external magnetic field. Based on our recent identification of spin polaron quasiparticle in Mott insulator \cite{davydova2022itinerant}, in this work we predict a new metallic state emerges at small doping and intermediate field range, a pseudogap metal that exhibits a single-particle gap and a doping-dependent magnetization plateau.
\end{abstract}

\maketitle

%\section{Introduction}

As the paradigmatic model for strongly correlated electron systems, the Hubbard model captures in the simplest form the essence of numerous electronic phenomena \cite{arovas2022hubbard}, including metal-insulator transition, metallic ferromagnetism, charge/spin stripe states and unconventional superconductivity. The Hubbard model was studied intensively in the context of cuprate high temperature superconductors \cite{lee2006doping}. Recently,  transition metal dichalcogenide (TMD) moir\'e heterostructures have emerged as a robust and tunable platform for the realization of Hubbard model physics \cite{PhysRevLett.121.026402,PhysRevB.102.201115,regan2020mott,tang2019wse2,li2021continuous,https://doi.org/10.48550/arxiv.2202.02055}. The moir\'e superlattice due to lattice mismatch or rotational twist introduces a long-wavelength periodic potential for itinerant charge carriers. At large moir\'e period, the system is akin to an array of artificial atoms, where each ``atom'' corresponds to a local minimum of the moir\'e potential and neighboring atoms are weakly coupled by tunneling through the potential barrier. As a result, narrow moir\'e bands are formed, and the band dispersion is well described by a tight binding model on the triangular or honeycomb lattice \cite{PhysRevLett.121.026402,PhysRevB.102.201115}. By further including Coulomb repulsion, a Hubbard model with on-site and non-local interactions \cite{PhysRevLett.128.217202} is obtained as the effective Hamiltonian of TMD moir\'e superlattices.

Compared to other Hubbard model materials, TMD moir\'e superlattice has a distinct advantage due to its robustness and tunability. The formation of narrow moir\'e bands needed for Mott-Hubbard physics does not require a magic twist angle. The band filling can be tuned continuously by electrostatic gating. The non-local interaction in moir\'e Hubbard systems can be screened by metallic gates \cite{tang2022frustrated}. The moir\'e bandwidth can be tuned by the twist angle and out-of-plane electric field \cite{li2021continuous,zhang2021spin}.

As a hallmark of Hubbard model physics at strong coupling ($U\gg t$), Mott insulating states have been observed in various TMD bilayers such as WSe$_2$/WS$_2$ at the filling of $n=1$ electron (or hole) per moir\'e unit cell \cite{regan2020mott,tang2019wse2}. The existence of local magnetic moments in the Mott state is directly revealed by magnetic circular dichroism (MCD) \cite{tang2019wse2,https://doi.org/10.48550/arxiv.2202.02055,tang2022frustrated}. At low magnetic fields, the temperature dependence of the MCD signal shows Curie-Weiss behavior with a negative Weiss constant that indicates antiferromagnetic exchange interaction $J=4t^2/U>0$. At low temperature, the MCD signal saturates at a field $B^*$ where the Zeeman energy $g\mu_B B^*$ becomes comparable to $J$.

%Because of the small kinetic energy scale $t\ll U$, complete spin polarization can be attained at a modest magnetic field $B^*$.

%In addition to Mott insulators, a variety of generalized Wigner crystals have been found in WSe$_2$/WS$_2$ at commensurate fractional fillings, which result from extended interactions on the moir\'e superlattice.
%Another remarkable discovery is the quantum anomalous Hall insulator in AB-stacked MoTe$_2$/WSe$_2$, which is a consequence of the inverted charge transfer gap in a two-band Hubbard model.

Up to now, theoretical studies of TMD moir\'e materials based on Hubbard model have mainly focused on the insulating states \cite{slagle2020charge,pan2020band,padhi2021generalized,xian2021realization,devakul2021magic,morales2021metal,devakul2022quantum,zang2022dynamical}. Much less explored are the metallic states that appear ubiquitously at generic filling fractions. Kinetic magnetism in the metallic states is beginning to be studied \cite{davydova2022itinerant,lee2022triangular}. Evidence of heavy Fermi liquids is being observed in MoTe$_2$/WSe$_2 $\cite{li2021continuous}. An intriguing open question is whether moir\'e Hubbard systems host metallic states that are fundamentally distinct from Fermi liquids.
%Signs of superconductivity have been reported, but remain to be confirmed.
%It is also unknown how non-Fermi liquids, if they exist, may be experimentally detected.

%strong interaction effects in the metallic states at or close to the filling of $n=1$ hole per moir\'e unit cell. twisted bilayer WSe$_2$ and AA-stacked MoTe$_2$/WSe$_2$

In this work, we predict a pseudogap metal phase in doped moir\'e Mott insulator at $n<1$ on the triangular lattice.
This phase is realized under a certain range of magnetic (Zeeman) fields and evidenced by an intermediate magnetization plateau corresponding to a total spin $S$ that is set by the doping density
$\delta=1-n>0$:
\begin{eqnarray}
S_p  = N(1-3 \delta)/2,
\label{Eq:mag}
\end{eqnarray}
where $N$ is the number of unit cells.
Our pseudogap metal is a non-Fermi liquid distinct from an ordinary metal by the presence of an energy gap of adding or removing an electron. Therefore, photoemission and tunneling measurements will find a single-particle gap at the Fermi level despite that the state is conducting and compressible. The pseudogap metal contrasts sharply with the fully spin polarized state at higher fields which has total spin $S_m=N(1-\delta)/2$ and is a conventional Fermi liquid, as well as the zero-field state which is a metallic antiferromagnet with $120^\circ$ order.

\begin{figure*}[ht]
	\includegraphics[width= 1.\textwidth]{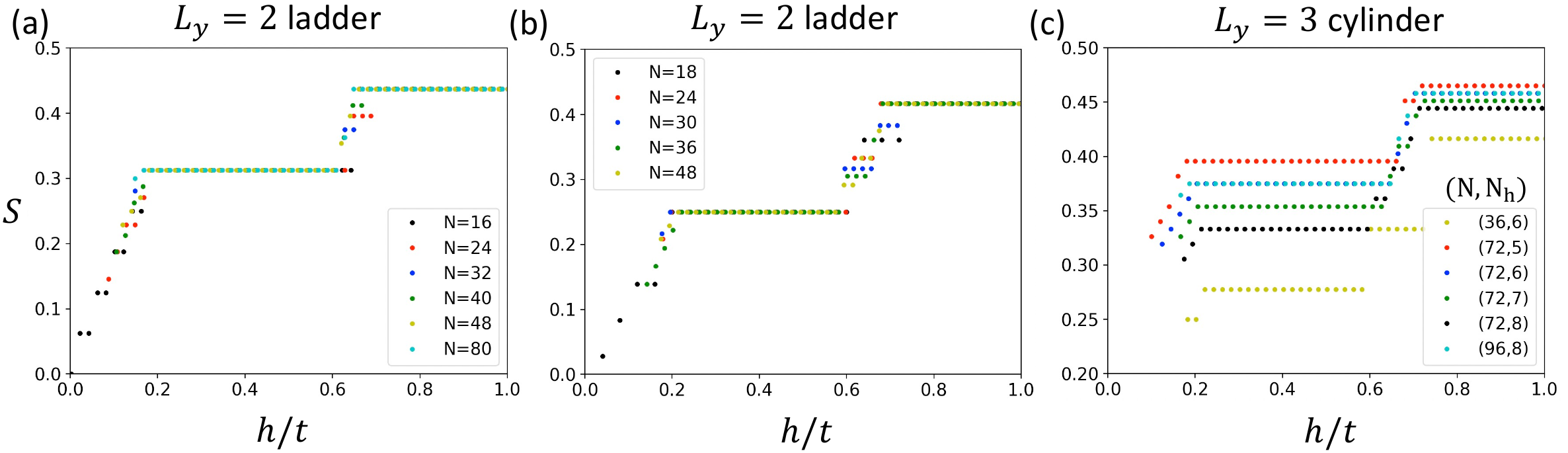}
	\caption{The total spin $S$ under magnetic field obtained by DMRG at infinite U limit. Finite-size scaling of (a) $\delta=1/8$ and (b)$\delta=1/6$ hole doping on two-leg ladder with $L_x$ up to 80. (c) $\delta$ dependent magnetization plateaus in $L_x=24,32$ three-leg cylinders. %We note the lines of $\delta=\frac{1}{12}$ from two cylinders overlaps. Our calculations stop after the left end of magnetization plateaus.
	} \label{fig:mp}
\end{figure*}

%Moreover, our exact diagonalization (ED) and  density matrix renormalization group (DMRG) results reveal a robust magnetization plateau under applied magnetic field at finite density of doped holes. The spin flips are tightly bounded to the doped holes from hole-spin correlation map.

% metallic and compressible  state.

%We also predict a half-metal phase in electron-doped Hubbard model on the honeycomb lattice with a staggered sublattice potential. This two-band Hubbard model captures the physics of charge-transfer insulators on the  moir\'e superlattice, where $\delta=n-1$ doped electrons occupy the $B$ sublattice to avoid double occupancy on the $A$ sites. This metallic phase also shows a fixed spin polarization
%\begin{eqnarray}
%s(n>1)= \frac{1-\delta}{1+\delta}. %\label{mag}
%\end{eqnarray}
%The defining feature of this half-metal is an energy gap to adding or removing an electron of the majority spin, but no gap for adding or removing an electron of the minority spin. In other words, our half-metal phase will have a spin-selective single-particle gap at the Fermi level as observed by tunneling or photoemission.

The microscopic origin of the pseudogap metal at $n=1-\delta$ can be traced to the nature of charge $-e$ quasiparticle in the Mott insulator at $n=1$. For the simplicity of notation, we consider electron (with charge $e$) filling of the moir\'e conduction band, so that $n<1$ corresponds to hole doping of the moir\'e Mott insulator.
As we showed in recent work with Davydova \cite{davydova2022itinerant}, for the triangular lattice Hubbard model in the strong-coupling regime $U\gg t$,  in a wide range of  magnetic fields,
the undoped Mott insulator is fully spin polarized, but the ground state with one doped hole is not, but contains one spin-flip that is bound to the hole \cite{zhang2018pairing}, resulting in an itinerant spin polaron that lowers the total energy. Remarkably, the formation of spin polaron has a kinetic origin associated with the correlated motion of the hole and spin-flip. The binding energy between the hole and spin-flip depends on the center of mass momentum $\bm k$ and reaches the maximum at ${\bm k}=0$, which is on the order of the hopping amplitude $t$. In the limit $U \rightarrow \infty$, we found by exact solution $\epsilon_b({\bm k}=0) \approx 0.42 t$ (also obtained in Ref.\cite{zhang2018pairing}). Therefore, for intermediate magnetic fields $h_1<h<h_2$ with $h_1 \sim J $ and $h_2 \sim t$, the low-energy charge $-e$ quasiparticle of the spin-polarized Mott insulator is the spin polaron instead of the bare hole.

Importantly, while bare hole has spin $s= \frac{1}{2}$ relative to the undoped and fully polarized Mott insulator, spin polaron has $s= \frac{3}{2}$ due to the extra spin-flip it carries.  Due to the difference in their spin quantum numbers, these two types of quasiparticles can be experimentally distinguished by measuring the lower Hubbard band edge as a function of the magnetic field \cite{davydova2022itinerant}. Following our theoretical prediction, a recent compressibility measurement reported evidence of a transition from spin polaron to bare hole excitation as $h$ increases. % above a critical field that corresponds to $h_2$.

This work is concerned with various types of metallic states at finite hole density ($n=1-\delta$) that arise under the application of a magnetic field. Of particular interest is that for small hole doping and at intermediate magnetic fields, a dilute gas of itinerant spin polarons may be formed, leading to an unconventional metal \cite{davydova2022itinerant}. From the fact that a spin polaron carries spin $s=\frac{3}{2}$, it immediately follows that  the total spin of the spin polaron metal $S_p$ is {\it locked} to the doping level as given by Eq.~(\ref{Eq:mag}), thus leading to a magnetization plateau with incomplete spin polarization.
It also follows that there is an energy gap to adding or removing spin-$\frac{1}{2}$ electrons, which are high-energy excitations orthogonal to the underlying spin-$\frac{3}{2}$ spin polarons. Thus, our spin polaron metal is a pseudogap metal distinct from Fermi liquids.
This picture is supported by our theoretical and numerical studies to be presented below.

%The situation is different at doping $n>1$. For the single-band Hubbard model on the triangular lattice,  the kinetic motion of doublons favors Nagaoka ferromagnetism, and the fully polarized state has a total spin  of $N(1-\delta)/2$ ($N$ is the total number of unit cells).
%In contrast, for two-band Hubbard model on the honeycomb lattice,
%we show that electron doping leads to the formation of \cmt{Zhang-Rice singlet} at low fields, while
%the high-spin state is only realized at high fields. While the fully polarized state has a total spin of $N(1+\delta)/2$, the Fermi liquid of \cmt{Zhang-Rice singlet} has a total spin of $N(1-\delta)/2$, hence the spin polarization given by Eq.(\ref{mag}).

%Strange metallic behavior has been observed in twisted AA-homobilayer WSe$_2$.

%\textit{Strong-coupling Hubbard model and the formation of spin polarons.---}
We study the triangular lattice Hubbard model for TMD heterobilayers  at electron filling $n \le 1$ ($n$ is the number of electrons per unit cell):   %we start with the Hubbard model on a triangular lattice with nearest-neighbour hopping $t$:
\begin{eqnarray}
H = -t \sum_{\langle i, j\rangle}\left(c_{i\sigma}^{\dagger} c_{j\sigma}+h . c .\right)+U \sum_{i} n_{i \uparrow} n_{i \downarrow} + \frac{h (N_{\uparrow}-N_{\downarrow})}{2}  \nonumber
\end{eqnarray}
where $c_{i \sigma}^{\dagger}(c_{i \sigma})$ is the fermion creation (annihilation) operator for spin $\sigma$ on site $i$, $n_{i \sigma}=c_{i \sigma}^{\dagger} c_{i \sigma}$ is the number operator and $h$ is the external magnetic field that couples to the $z$-component of the total spin $S_z$.
We consider the case of large moir\'e wavelength, where the kinetic energy $t$ is much smaller than the onsite repulsion $U$.
%Non-local interactions are also parametrically smaller than $U$ and their effect  will be discussed later.

At half-filling ($n=1$), the ground state is an antiferromagnetic Mott insulator whose magnetic property is governed by the Heisenberg model on the triangular lattice: $H_{J}=J \sum_{\langle i j\rangle} \boldsymbol{s}_{i} \cdot \boldsymbol{s}_{j}$. In the experimentally relevant strong-coupling regime $U\gg t$, the exchange interaction $J$ is very small and the Mott insulator becomes fully polarized above a small magnetic field $h_1=\frac{9}{2}J$, which is on the order of $1$ T for WSe$_2$/WS$_2$ \cite{tang2019wse2}.

In contrast, at finite doping, magnetism arises predominantly from the kinetic motion of doped charges, with an energy scale set by the hopping amplitude $t$ that is much larger than $J=4t^2/U$. To highlight the kinetic mechanism for magnetism, we shall mainly focus on the Hubbard model in the infinite-$U$ limit.
\begin{figure}[htp]
	\includegraphics[width= 1.0\columnwidth]{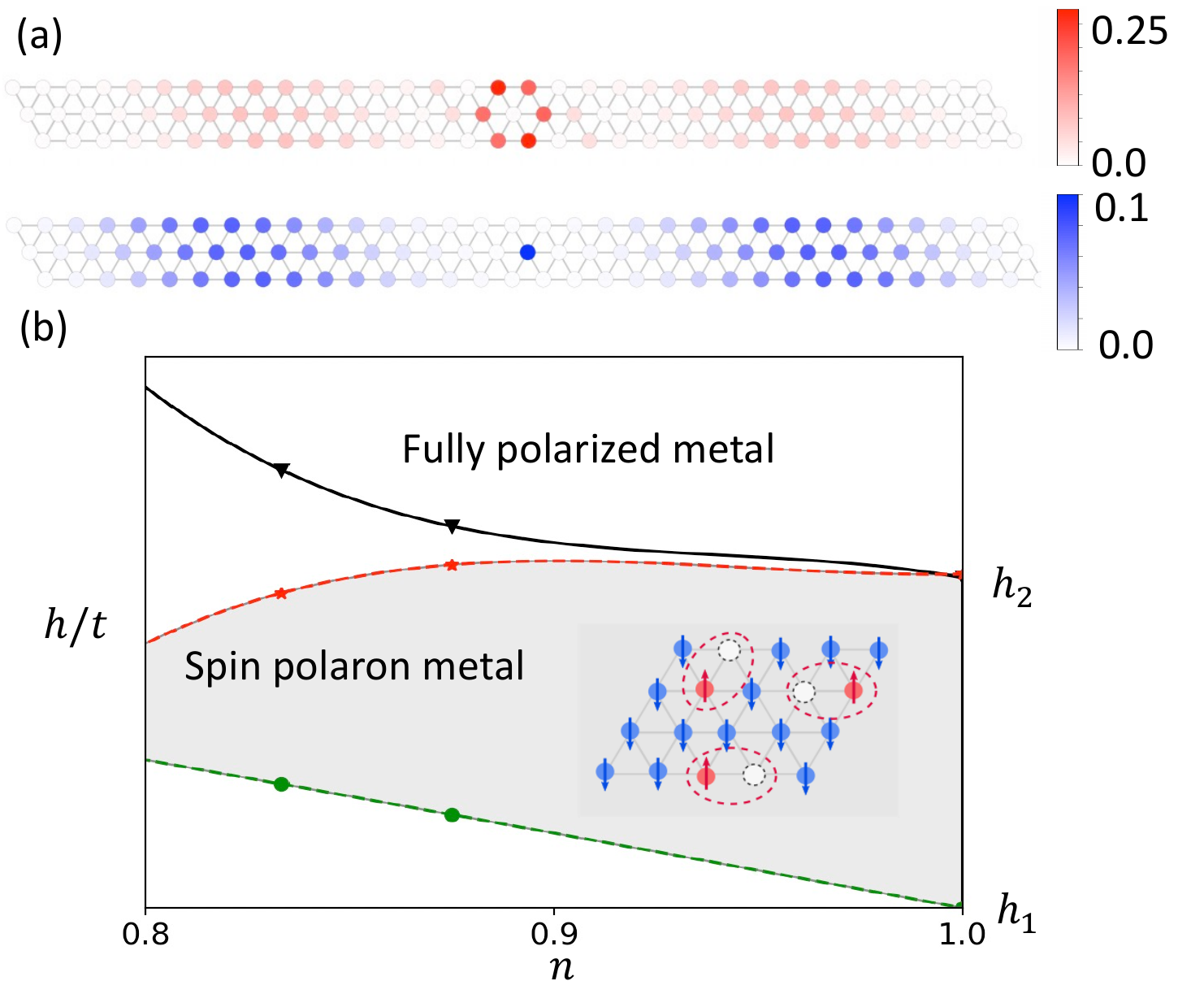}
	\caption{(a) Hole-spin correlation function $C_{hs} (i, o)/\delta$ in a $33\times 3$ cylinder with 3 spin polarons, where holes are tightly bounded to the spin flips. The bottom panel is the Hole-Hole correlation function $C_{hh} (i, o)/\delta$. $o$ is the center of cylinder. (b) Schematic figure of doping- and field-dependent phase diagram. The data points are taken from DMRG on a $24\times3$ cylinder.
	} \label{fig:sche}
\end{figure}

%In previous work, we have shown that at single hole doping, the charge excitation is a spin polaron when the magnetic field is reduced slightly under the saturation field. The binding energy of a single spin polaron is purely kinetic origin, which induces the different fully polarization field at electron and hole doped side in triangular lattice. In this work, we will study the nature of dilute gas of spin polarons, and its behavior under magnetic field.

We use exact diagonalization and density matrix renormalization group (DMRG) methods \cite{white1992density} to study the ground state of $H$ as a function of magnetic field and doping.
ED calculation is performed on two-leg ladders ($L_y$=2) with $L_x$ between 8 and 21. To reduce the finite-size effects, periodic and anti-periodic boundary conditions in $x$ direction are used for even and odd number of holes, respectively. Since particle number and total spin $S_z$ are conserved, we divide the full Hilbert space into $(N_\uparrow, N_\downarrow)$ sectors to reduce the computational cost.

%The full Hilbert space is then divided into spin and momentum sectors to approach the large system size. We start with the two-leg ladder geometry
%\new{ From the  tight-binding calculation in a sector with one hole and one spin flip (see \cite{} and  SM for the details),  we obtain that the binding energy between a hole and spin flip for a single spin polaron on a 2-leg ladder  approaches $\varepsilon_b^{(1)} \approx 0.606 t$ in the limit $L_x \rightarrow \infty$. }

We further perform quantum number conserving density renormalization group (DMRG) calculation \cite{white1992density,zaletel2015infinite,jiang2008density}, as implemented in the ITensor package\cite{itensor}, for two-leg ladders and multi-leg cylinders ($L_y=3, 4$ and $6$) with open boundary condition along the $x$ direction, reaching system sizes up to $40\times 2$, $32\times 3$ and $12\times 6$. %Optimal convergence is attained by increasing the number of bond dimensions for matrix product states.
The convergence of our DMRG calculation %of the exact diagonalization step
is improved by keeping track of the basis transformations and using them to construct a good initial guess for the next step wavefunctions. We introduce a random noise of $10^{-6}$ to $10^{-8}$ at first few steps to avoid the local minimum trapping. The maximum bond dimension and cutoff is set to be 50000 and $10^{-7}$, the convergence criteria is $10^{-7}$ of the total energy. As a benchmark, we compare the ground state property of two-leg ladders with ED and find excellent agreement.  %up to $10^{-7}$. %The density distribution and magnetization curve obtained from ED and DMRG are found to be nearly identical.
%We utilize $2\times L_x$ ($L_x$ up to 40) ladders and $3\times L_x$ ($L_x$ up to 32) cylinders and calculation hole doping up to $\delta=\frac{1}{6}$ around half-filling. The site index are chosen to increase along the narrow $y$ direction, as shown in the Fig. S2.

%Without spin coupling, the $tJ$ and canonical Hubbard Hamiltonian have $SU(2)$ symmetry, and the DMRG calculation preserves the quantum numbers during the convergence steps. Consequently, we parallelize the magnetic plateau calculations over different particle number and spin sectors.

%\textbf{Magnetization plateaus}
Fig. \ref{fig:mp} shows the ground state magnetization (total spin $S$) at various hole doping densities as a function of magnetic field $h$. For the two-leg ladder at $\frac{1}{8}$ and $\frac{1}{6}$ dopings, magnetization curves $S(h)$ clearly converge to the thermodynamic limit with increasing system sizes $L_x$. % and calculate the field-dependent magnetization at zero temperature.
%At $h=0$, the ground state is
Full spin polarization is attained at high fields $h>h_2 \sim t$, showing that kinetic energy of holes governs magnetic properties in hole-doped Mott insulator at large $U/t$. As the magnetic field is reduced, a magnetization plateau is observed over an intermediate range of fields, with the corresponding total spin $S_p$ equal to Eq.(\ref{Eq:mag}). This is exactly expected from the spin polaron picture: every doped hole is bound to a single spin-flip, so that the number of spin flips is equal to the number of doped holes.

As shown in Fig.\ref{fig:mp}(c), magnetization plateau is also found in three-leg cylinders for various hole dopings up to at least $\delta= \frac{1}{6}$. For $\delta=\frac{1}{12}$, the magnetization curves at two different system sizes $L_x=24, 36$ nearly coincide, showing the convergence to the thermodynamic limit. As in the two-leg case, the spin polarization on the magnetization plateau $S_p$ as a function of doping exactly matches the formula Eq.(\ref{Eq:mag}) expected for the spin polaron metal. With increasing hole density, the width of magnetization plateaus shrinks as interaction effect between spin polarons becomes important.

The presence of spin polaron  is further supported by the spatial correlation between hole and spin-flip in the ground state.  We calculate the %density correlation  $\la n_h(i)n_h(o) \ra$ and
real-space correlation function between hole and minority spin (spin-flip):
%\begin{eqnarray}
$C_{hs} (i, j) = \la   n_h(i)  n_{\downarrow}(j) \ra$,
%\end{eqnarray}
where $n_h(i)= 1-\sum_{\sigma} c_{i \sigma}^{\dagger} c_{i \sigma}$.  %and $n_{\downarrow}(i)= c_{i \downarrow}^{\dagger} c_{i \downarrow}$.
As shown in Fig. \ref{fig:sche} for three-leg cylinder, in the presence of a spin-flip, there is a very high probability ($>72\%$) of finding a hole on its nearest-neighbor sites, indicating a tightly bound state of hole and spin-flip.
We also calculate the density correlation $C_{hh} (i, j) = \la   n_h(i)  n_h(j) \ra$ and find doped holes stay away from each other,  indicating the repulsive interaction between spin polarons. These correlation functions also show that at small doping, each hole spreads over many sites, consistent with its itinerant character.

Our recent theoretical study \cite{davydova2022itinerant} shows that the energy dispersion of a single spin polaron has a unique minimum at $\Gamma$ with an effective mass significantly larger than the mass of bare hole. %$m^*=\frac{26}{3 ta^2}$ ($a$ is lattice constant), which is about 13 times the mass of bare hole.
Then, at small hole doping, the dilute gas of spin polarons has a small Fermi surface centered at $\Gamma$. In contrast, the fully polarized state at high fields  is a dilute gas of bare holes, which has two disconnected Fermi pockets around $K$ and $K'$ valley. Therefore, for a given hole density, the spin polaron metal and fully polarized metal, which appear at different ranges of magnetic field, have distinct Fermi surfaces with different Fermi wavevectors.

\begin{figure}[h]
	\includegraphics[width= 1.\columnwidth]{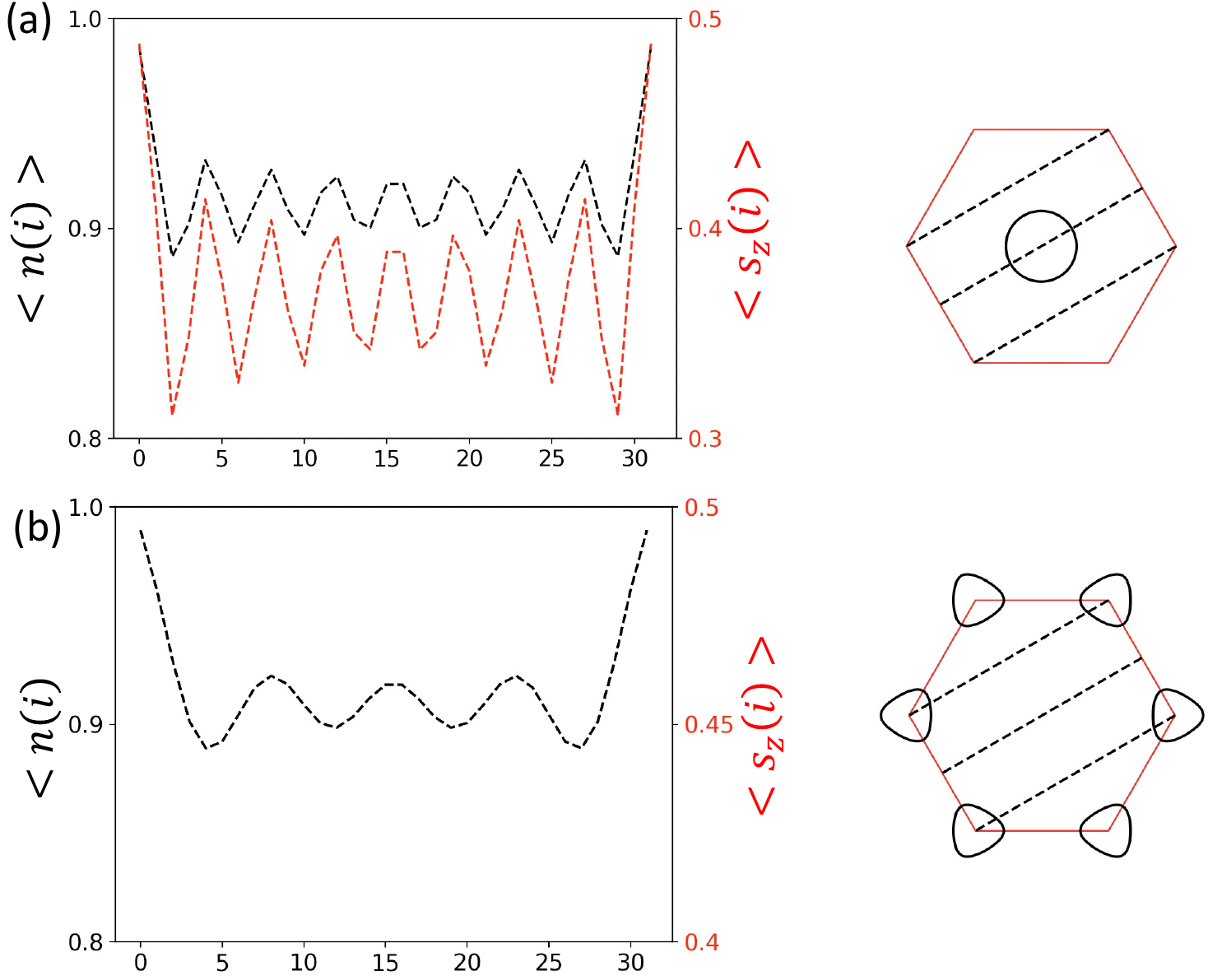}
	\caption{(a) Friedel oscillation of charge and spin densities in $32\times3$ cylinder with $\delta=\frac{1}{12}$ hole doping in spin polaron metal; (b) Friedel oscillation in fully polarized metal of the same system.
    The two metallic states at the same density have different Fermi wavevectors, leading to different oscillation periods. }
	\label{fig:oss}
\end{figure}

We now demonstrate the existence of Fermi surface at small doping by observing Friedel oscillations of charge and spin densities in the presence of boundaries. %In one-dimensional and quasi-one-dimensional systems,
The period of Friedel oscillation is given by $2\pi/2k_F$,
%decays as $\delta \rho(x) \sim \cos \left(2 k_{F} x+\varphi\right)/x^{g}$,
where $k_F$ is the Fermi wavevector. Fig.(\ref{fig:oss}) shows density distributions in real space at $\delta=\frac{1}{12}$ hole doping for $32 \times 3$ cylinder with open boundary condition in the $x$ direction.
Pronounced Friedel oscillations are observed in both spin polaron metal and fully polarized metal. The periodicity of the oscillation is $4a$ in spin polaron metal, and $8a$ in fully polarized metal.

These results can be understood straightforwardly by taking into account of the finite circumference of the cylinder. In this geometry, the 2D Fermi sea is sliced into a number of 1D Fermi sea at various $k_y=2\pi n /L_y$ with $n=0, ..., L_y-1$.
For $L_y=3$ and $\delta=\frac{1}{12}$, in spin polaron metal, the line $k_y=0$ cuts through
a single Fermi surface centered at $\Gamma$. The corresponding $2k_F$ leads to an oscillation period $a/(\delta L_y)=4a$. In contrast, in fully polarized metal, the Fermi wavevector is halved due to the presence of two disconnected Fermi surfaces, leading to a doubled period of $8a$.

%\textbf{Fermi surface of $6\times 9$ cluster}

%\textbf{Compressibility based on ground state energy as a function of doping}

Our results so far have identified an unconventional metallic phase---a dilute gas of spin polarons---in the triangular lattice Hubbard model at small hole doping and intermediate magnetic field. A hallmark of spin polaron metal is that its zero-temperature magnetization is determined solely by the doping level as given by Eq.{\ref{Eq:mag}}. It is a compressible state with a single Fermi surface centered around $\Gamma$.

We now demonstrate another defining feature of spin polaron metal: it has a single-particle gap, i.e., adding/removing an electron costs finite energy. Intuitively, the reason is obvious: the constituent particles forming the metallic state are spin polarons carrying spin-$\frac{3}{2}$, while the electron carrying spin-$\frac{1}{2}$ is a high-energy excitation. To show the single-particle gap explicitly, let us denote the ground state energy at a given total particle number $N$ and total spin $S$ by $E(N, S)=E_0(N, S) - h S$, where $E_0$ is the ground state energy at $h=0$. Minimizing $E(N,S)$ over all possible values of $S=0, ..., N/2$ yields the $N$-particle ground state energy $E_N \equiv \min_{S} E(N,S)$ as well as the corresponding spin which is denoted as $S_N$. Adding an electron necessarily increases or decreases the total spin by $\frac{1}{2}$. The single-particle gap is thus defined by
\begin{eqnarray}
\Delta E_{e, s} = E(N + 1, S_N + s) - E_N - \mu,
\end{eqnarray}
with $s=\pm \frac{1}{2}$ depending on the added electron being spin $\uparrow$ or $\downarrow$. $\mu$ is the chemical potential defined by $\mu = \frac{\partial E_N}{\partial N}$.
%In spin polaron metal phase, we have
For metallic states in the thermodynamic limit $N\rightarrow \infty$, $\mu = E_{N + 1} - E_N$. %or $E(N) - E(N-1, S_N - \frac{3}{2})$.
Then, the single-particle gap is equal to
$\Delta E_{e, s} = E(N, S_{N-1} + s) - E_N$.
In the spin polaron metal phase, we have $S_{N-1} = S_{N} - s_0 $ with $s_0=\frac{3}{2}$ because the $(N-1)$-particle state has one more spin polaron than the $N$-particle state.
It thus follows that the single-particle gap is equal to the spin gap, i.e.,  the energy cost resulting from the inevitable spin mismatch $s -\frac{3}{2} \neq 0$ between spin polaron and the added electron:
\begin{eqnarray}
\Delta E_{e, s} = E(N, S_{N}  + s - s_0) - E_N.
\end{eqnarray}
%Depending on the direction of the spin of the added electron, $s-\frac{3}{2}$ is either $-1$ or $-2$.
Importantly, the presence of magnetization plateau over a finite range of magnetic fields implies the existence of spin gap in the spin polaron metal (otherwise the total spin would change continuously with $h$). We thus conclude that spin polaron metal has  a single particle gap. It should be contrasted with the fully polarized metal, which also has a spin gap. In that case, $s_0=\frac{1}{2}$ hence $\Delta E_{e, +\frac{1}{2}}=0$, i.e.,  there is no gap to adding an electron of spin $\uparrow$.

%whereas in fully polarized metal, two lines $k_y=\pm 2\pi/3$ cut through the Fermi leading to Friedel oscillation shown in Fig.(\ref{fig:oss}).     On the magnetization plateau, the
%The pronounced Friedel oscillation of electron density indicates the metallic nature of spin polaron metal.

To summarize, the main finding of our work is spin polaron metal in doped Mott insulator on the triangular lattice under a magnetic field. It is a remarkable state of matter at generic fillings, which is conducting and compressible similar to an ordinary metal, but has a spin gap and a single-particle gap. Its existence can be can experimentally established by a zero-temperature magnetization $S_p$ that depends only on the doping as given by Eq.(\ref{Eq:mag}). Increasing the magnetic field drives a transition from the spin polaron metal to the fully polarized metal, accompanied by a change of Fermi surface volume that can be detected by the change of Landau level degeneracy and quantum oscillation frequency.

Finally, we briefly discuss the ground state of infinite-$U$ Hubbard model at smaller magnetic fields $h<h_1$. As $h$ is reduced, the magnetization decreases continuously to zero at $h=0$. For three-leg cylinders, we do not observe any additional plateau that could be associated with bound state between hole and multiple spin-flips \cite{zhang2018pairing,morera2021attraction}. At zero field, we find that small doping induces strong antiferromagnetic correlation wavector $K, K'$ consistent with $120^\circ$ order,  consistent with a recent study \cite{zhu2022doped}. In the magnetization curves shown in Fig. \ref{fig:mp}, the magnetization plateau stands out as the most prominent feature of doped Mott insulator on the triangular lattice, which we identify as the hallmark of a pseudogap metal composed of itinerant spin polarons.

%an unconventional metallic state

\section*{Acknowledgments}
We thank Margarita Davydova for collaboration on a recent work \cite{davydova2022itinerant} which led to this study.

%\section{Half metal at $n>1$}
%Beyond the triangular lattice Hubbard model, we now discuss the behavior of the charge transfer insulator at the electron doping side, $n>1$. In the recent experiment of twisted homobilayer WSe$_2$ with AB stacking, it was found that the charge carriers become fully polarized even for $n>1$ at the electron doping side, indicating the existence of a second moir\'e region. To accurately extract the moir\'e potential landscape, we perform large-scale density functional theory calculation to determine the electronic structure of the supercell.

%\vfill
%\clearpage

\bibliography{ref}

\clearpage
\pagebreak
\onecolumngrid
\begin{center}
\textbf{\large Supplemental Materials: Pseudogap metal and magnetization plateau from doping moir\'e Mott insulator}
\end{center}

%\subsection{Tight-binding calculation for a single spin polaron on a 2-leg ladder}
%\cmt{[a plot and discussion will be added]}

\subsection{Quantum number conserving Density Matrix Renormalization Group}
In this section, we present the detailed description of our DMRG calculation. We employ the DMRG algorithm as implemented in ITensor package\cite{itensor} on the finite ladder and cylinder. And the $x$ direction is chosen as the long direction with open boundary condition to reduce entanglement and bond dimension. We utilize $2\times L_x$ ($L_x$ up to 40) ladder and $3\times L_x$ ($L_x$ up to 32) cylinder and calculation hole doping up to $\frac{1}{6}$ around half-filling. The site index is chosen to increase along the narrow $y$ direction. For small system size (site number less than 40), a random matrix product state (MPS) with fixed number of holes $N=N_{\uparrow}+N_{\downarrow}$ and spin flips $N_{\downarrow}$ is generated as the initial wavefunction.
Without spin coupling, the $tJ$ and full Hubbard Hamiltonian have $SU(2)$ symmetry, and the DMRG calculation preserves the quantum numbers during the convergence steps. Consequently, we parallelize the magnetic plateau calculations over different particle number and spin sectors.

When the initial MPS is close to the global minimum,
robust convergence can be achieved with a few DMRG steps and moderate bond dimensions. However, if the initial MPS is far from the global minumum then there is no guarantee that DMRG will be able to find the true ground state. In our problem, occupation and spin quantum number are conserved to reduce the Hilbert space dimension. The convergence problem is much more significant since the search space is more constrained.

To improve the numerical stability, here we add noise term from $10^{-6}$ to $10^{-10}$ to avoid local mimimum trapping effect. More importantly, we create the initial MPS with properties close to the ground state in terms of charge configuration and spin ordering. Especially for the case of multi spin polarons at two or three-leg ladders, we can approximate the trial MPS of large system from the combination of converged MPS of a small system.
Therefore, the electron density is spread out over the whole system and the spin flips are binded to the hole, as expected for the final MPS. The creation of initial MPS greatly reduces the computation cost and bond dimension for large system size, and allows the computation for 16 holes in a $3\times 32$ cylinder, which has a Hilbert space dimension 1.7*$10^{34}$ at the sector with $N_{\downarrow}=16$ in the $tJ$ basis.

For the accurate evaluation of magnetic plateau and correlation function, we set energy convergence criteria as $10^{-7}t$, the cutoff as $10^{-8}t$, and maximum bond dimension as 40000. We will compare the energy with exact diagonalization for the case with small Hilbert space in the next section and the energy difference is normally within $10^{-8}t$. With the converged ground state MPS, we calculate the electron density distribution, hole-hole $\la n(0)n(x) \ra$ and hole-spin $\la n(0)s_z(x) \ra$ correlations in real space, where we defined $n(i)= 1-\sum_{\sigma} c_{i \sigma}^{\dagger} c_{i \sigma}$ as the hole density operator.

%For the 2D cylinder simulation, we further consider the 4 leg and 6 leg supercells to check the width of corresponding magnetization plateaus.
\begin{figure}[h]
	\includegraphics[width= 0.8\columnwidth]{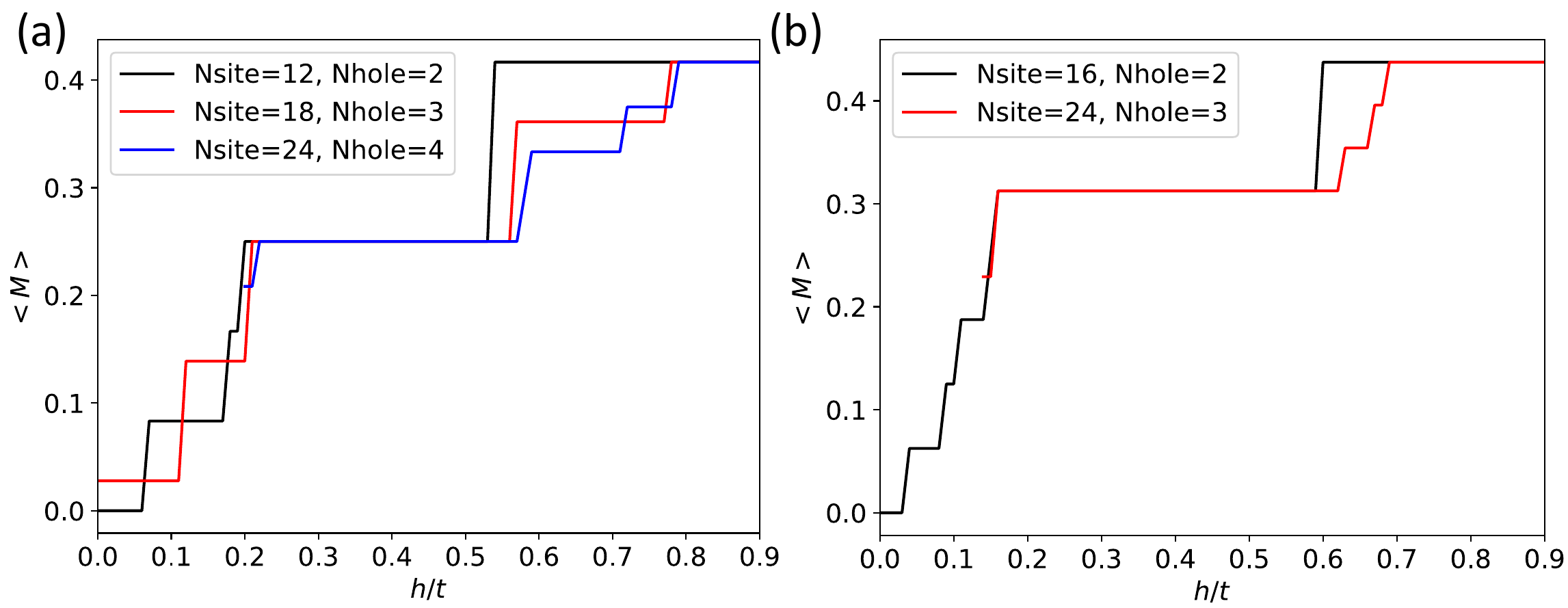}
	\caption{The magnetic moment $M$ under magnetic field obtained by exact diagonalization at infinite U limit and $1/6$ and $1/8$ hole doping on two-leg ladder with $L_x=6, 9, 12$ and periodic boundary condition in $x$ direction.
	} \label{fig:mp-ed}
\end{figure}

\textbf{Comparison with exact diagonalization.} We compare the ground state energy from DMRG with exact diagonalization. At a computing node with $500 GB$ memory, the up limit for Hilbert space dimension in exact diagonalization is $2 \times 10^{9}$ if we use float128 to represent the basis and Hamiltonian. Within the $tJ$ model at $J=0$ limit, we compare the ground state energy, real space electron and spin density for 4 holes at $2\times 12$ ladder. As shown in Fig. \ref{fig:mp-ed}, the magnetization plateaus and real space electron density ($N_{\downarrow}=4$) plots from ED and DMRG almost overlap with each other.

\subsection{Antiferromagetism at finite doping}
We study the ground-state properties of infinite U Hubbard model at total spin $S^z=0$ spin sector \cite{zhu2022doped}. At filling factor $n=1$, the spin exchange $J=4\frac{t^2}{U}$ is vanishing small at infinite U limit. At finite hole doping side, we now consider antiferromagnetic interaction at zero magnetic field. In the ultra-strong coupling $U \gg t$ limit, spin correlation arises from kinetic term. And we probe the ground state magnetic property of the $J=0$ $tJ$ model by calculating the spin structure factor $S_{\mathbf{q}}(\mathbf{Q})=\frac{1}{N} \sum_{i, j}\left\langle S_{i}^{z} S_{j}^{z}\right\rangle e^{i \mathbf{Q} \cdot\left(\mathbf{r}_{\mathbf{i}}-\mathbf{r}_{\mathbf{j}}\right)}$. Here we consider the number of holes from 0 to 16 in a $3\times 24$ and $4\times 18$ cylinder, covering $0\%$ to $22\%$ doping density. At zero magnetic field, the total $\left\langle S_{i}^{z}\right\rangle$ is vanishingly small. In the DMRG calculation, we target at the spin sector with $\left\langle S_{i}^{z}\right\rangle=0$ by enforcing $N_{\uparrow}=N_{\downarrow}$.

\begin{figure}[h]
	\includegraphics[width= 0.6\columnwidth]{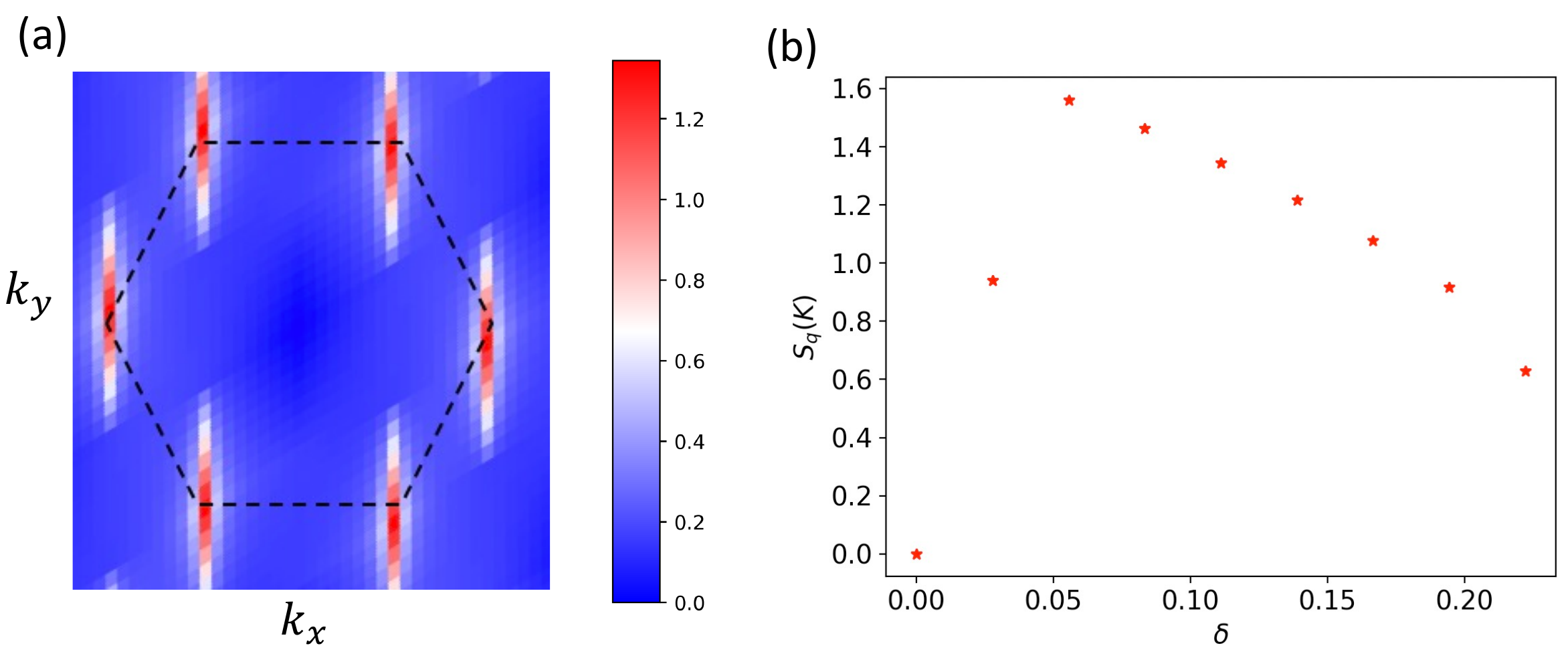}
	\caption{The static spin structure factor $S_q(Q)$ on $24\times 3$ cylinder for infinite $U$ Hubbard model. (a) Contour plot of $S_q(Q)$ at hole doping $\delta=5.5\%$, which shows a peak at momentum $Q=K$. (b) Maximum $S_q(Q)$ at momentum $K$ as a function of hole doping, indicating an enhanced antiferromagnetic exchange at finite doping density.
	} \label{afm}
\end{figure}

\textbf{Doping dependence of kinetic antiferromagnetic exchange} And we confirmed that $S_{\mathbf{q}}(\mathbf{Q})$ is zero at all momentum. By doping even number of holes into the $24\times 3$ ladder, we find the $S_{\mathbf{q}}(\mathbf{Q})$ becomes nonzero and sharply peaks at $\mathbf{Q})=\mathbf{K})$ momentum as shown in the contour plot in \ref{afm}(c), indicating the kinetic induced antiferromagnetism with $120^{\circ}$ noncollinear order. In \ref{afm}(b), we plot the maximum $S_{\mathbf{q}}(\mathbf{K}$ as a function of hole doping density. Starting from massive degenerate spin disorder phase at $n=1$, the antiferromagnetism gets greatly enhanced at light doping region with a peak at $\delta=5.5\%$, and suppressed with further enlarged hole doping.
\begin{figure}[h]
	\includegraphics[width= 0.8\columnwidth]{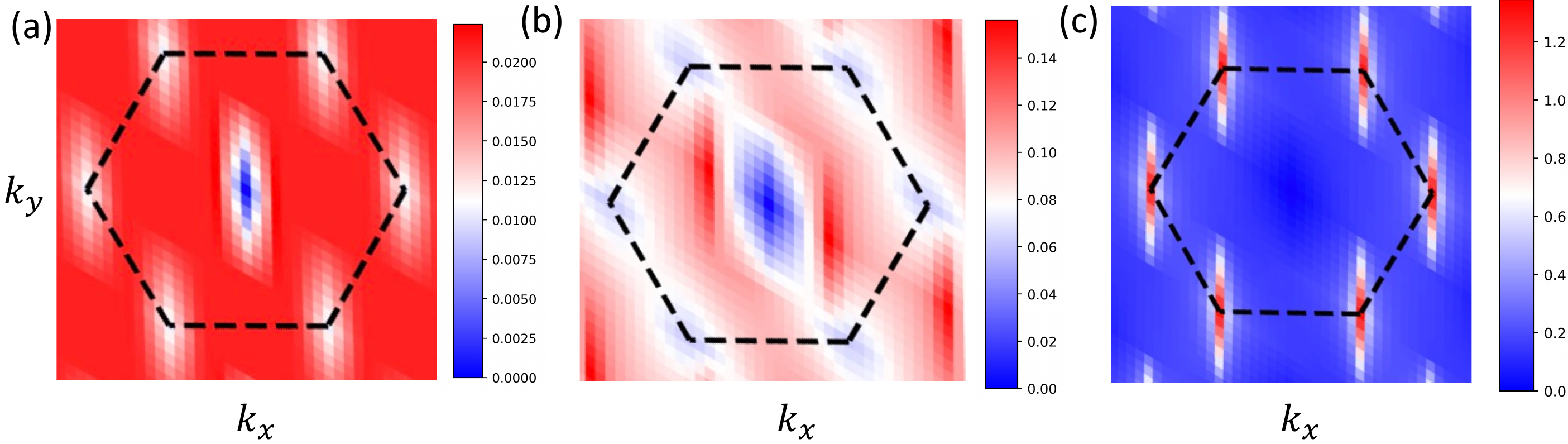}
	\caption{The static spin structure factor $S_q(Q)$ on $24\times 3$ cylinder for infinite $U$ Hubard model with hole doping $\delta=5.5\%$ (6 holes). (a) Fully polarized spin sector. (b) Spin polaron sector with 6 spin flips. (c) Antiferromagnetic spin sector with total $<S_z>=0$.
	} \label{afm-cor}
\end{figure}
We further calculate the spin exchange as a function of total magnetization $<S_z>$ in a $24\times 3$ cylinder. For the fully polarized spin sector, the spin structure factor map has deeps at $\Gamma$ and $K$ momentum. When the number of spin flip is equal to the number of holes, the spin polaron spin structure map has two sharp peaks around $\Gamma$ momentum as shown in Fig. \ref{afm-cor}.
%at $2k_F$ ($k_F$ is the wavevector of Fermi surface).

\subsection{Magnetization plateau in the finite-$U$ Hubbard model}
As described in the main text, the width of magnetization plateau stay unchanged when $U$ is reduced from infinite limit. We now present the detailed calculation of $U=10,20,40$ Hubbard model. In real material systems, the onsite repulsion $U$ is always finite. We consider the case of finite $U$, and the corresponding magnetization plateau within the %reduced tJ model and
Hubbard model. %After excluding the double occupancies in the original Hilbert space, the effective model exactly downfolds to the $t-J$ model with finite $J$.
%\begin{equation}
%\hat{H}=-t \sum_{\langle i j\rangle, \sigma}\left(c_{i \sigma}^{\dagger} c_{j \sigma}+\text { h.c. }\right)+ J \sum_{\langle i j\rangle}\left(\mathbf{s}_{i} \cdot \mathbf{s}_{j}-\frac{n_{i} n_{j}}{4}\right)
%\end{equation}
%Here $\mathbf{s}_{j}=\sum_{\sigma \sigma^{\prime}} c_{i \sigma} \frac{\boldsymbol{\sigma}_{\sigma \sigma^{\prime}}}{2} c_{i \sigma^{\prime}}$ is the spin operator. With ED, we calculate the magnetization vs. magnetic field for the spin exchange $J=0.1t$ from $U=40t$. As shown in Fig. \ref{fig:mp2}, the critical fields for fully polarized state increases, while the plateau width stays unchanged.

%\begin{figure}[h]
%	\includegraphics[width= 0.8\columnwidth]{Fig-MP-ED-J.pdf}
%	\caption{The magnetic moment $M$ under magnetic field obtained by exact diagonalization of the $t-J$ model with $J=0.1t$ and $1/6$ and $1/8$ hole doping on two-leg ladder with  $L_x=6, 9, 12$ and periodic boundary condition in $x$ direction.
%	} \label{fig:mp2}
%\end{figure}

%When downfolding from Hubbard model to $tJ$ model, an important term called correlated hoping is neglected. However, the correlated hopping plays an important role when the system is away from the strong coupling regime or at large doping density.
To properly include the effect of double occupancy at finite coupling strength, we directly diagonalize the Hubbard model with double occupancy allowed Hilbert space for U=$10, 20, 40$ using DMRG. For three-leg ladder with $L_x=24$ with finite $U$, the magnetization plateaus get shifted to higher magnetic field, which is consistent with enlarged saturation field\cite{davydova2022itinerant}. The width of magnetization plateau stay unchanged with decreasing onsite interaction down to $U=10t$. Therefore, we conclude the presence of spin polaron metal at wide range of onsite repulsion.
\begin{figure}[ht]
	\includegraphics[width= 1.0\textwidth]{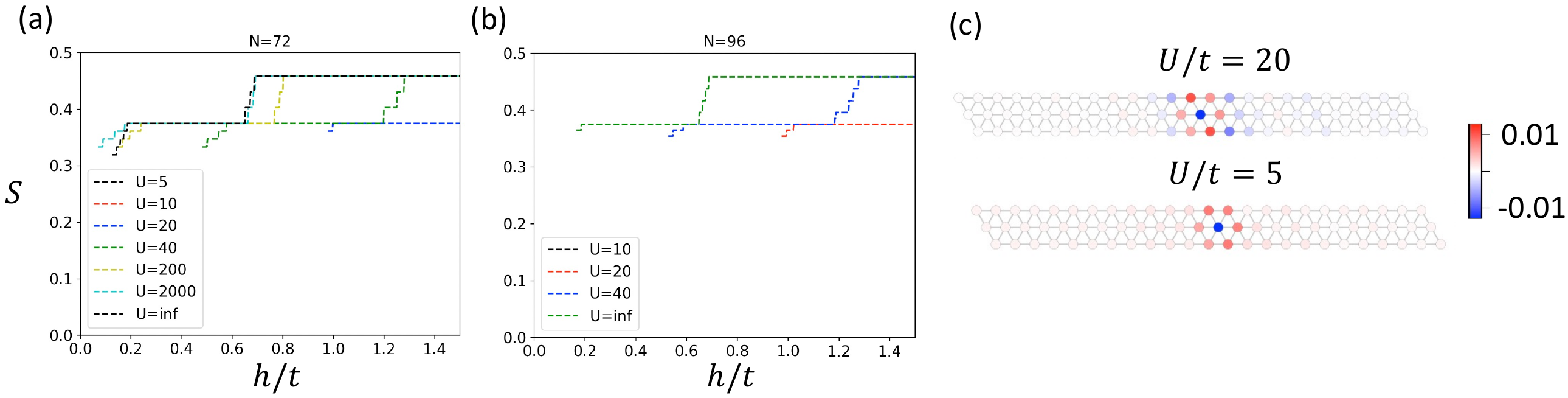}
	\caption{The total spin $S$ under magnetic field obtained by DMRG for different $U$ at (a) $\delta=1/12$ for $24\times 3$ cylinder,  (b)$\delta=1/12$ for $24\times 3$ cylinder. (c)
	Hole-minority spin correlation function normalized by hole density $C_{hs} (i, o)/\delta$ for 6 spin polarons in a $25\times 3$ cylinder. Upper and bottom panel for $U/t=5,20$, respectively.
	} \label{fig:Hubbard}
\end{figure}

%\textit{Pseudogap in the hole-doped spectrum.} In the angle-resolved photoemission (ARPES) experiment, the spectral function can be obtained by Fermi-function-division of the removal spectrum. And the pseudogap is defined as the suppression of electronic density of states near the Fermi level, most known in the hole-doped cuprates.

%Next, we calculate the zero-temperature momentum-resolved spectral function at a finite number of spin polarons:
%\begin{equation}
%C(\omega,q)=-\frac{1}{\pi}\left\langle 0\left|c_q^{\dagger} \frac{1}{\omega+i \eta+E_{0}-H} c_q\right| 0\right\rangle
%\end{equation}
%where $\eta=0.05t$ is the broadening factor for the spectral peaks, $\ket{0}$ is the ground state wavefunction, and $
%c_{q}=\frac{1}{\sqrt{L_y}} \sum_{r}^{L_y} \exp \left(-i \frac{2 \pi q r}{L_y}\right) c_{r}$. For an odd number of spin polarons, the ground state wavefunction is uniquely located in the zero-momentum sector. We first calculate the spectrum of one spin polaron at a $2\times 21$ ladder.

%\textbf{Entanglement entropy of Pseudogap metal.} The local entropy scaled with system size has been used to probe the correlated metallic phase in doped Mott insulator. Here we measure the entanglement entropy as a function of spin exchange interaction, to detect the breakdown of Pseudogap metal around $U\sim 15$.

\end{document}